\documentclass[11pt]{amsart}
\usepackage{latexsym,amssymb,amsmath,amscd,amsthm}
\usepackage{amsfonts}
\numberwithin{equation}{section}
\theoremstyle{remark}

\newcommand{\bq}{\begin{equation}}
\newcommand{\bea}{\begin{array}}
\newcommand{\eea}{\end{array}}

\newcommand{\ga}{\alpha}

\newcommand{\gD}{\Delta}
\newcommand{\gl}{\lambda}
\newcommand{\gL}{\Lambda}
\newcommand{\gb}{\beta}

\newcommand{\mf}{\mathfrak}

\newcommand{\gO}{\Omega}
\newcommand{\gG}{\Gamma}

\newcommand{\gs}{\sigma}

\newcommand{\gag}{\gamma}
\newcommand{\gd}{\delta}
\newcommand{\pp}{\partial}

\newcommand{\tl}{\tilde}
\newcommand{\na}{\nabla}

\newcommand{\bs}{\blacksquare}

\newcommand{\bgs}{\bigstar}

\newcommand{{\DDD}}{D\!\!\!\!\!\!-}

\title{REMARKS ON GRAVITY AND QUANTUM GEOMETRY}

\author{Robert Carroll\\University of Illinois, Urbana, IL 61801}

\date{July, 2010\thanks{email: rcarroll@math.uiuc.edu}}

\begin{document}


\bibliographystyle{plain}


\begin{abstract} 
Some relations between conformal relativity and Weyl-Dirac theory are
rephrased and clarified.
\end{abstract}

\maketitle


\section{BACKGROUND}
\renewcommand{\theequation}{1.\arabic{equation}}
\setcounter{equation}{0}

In recent times, stimulated by seminal paper of E. Verlinde \cite{vlde}, there has been
a deluge of material about entropy and gravity (we refer to Padmanabhan \cite{pada,padn,pdbh} for background and some references to the new material along with
some striking new results on equipartition).  The strong connections of 
thermodynamics, entropy, and gravity have been in the air for some time, going back
to Beckenstein, Jacobson, and Padmanabhan (cf. \cite{beck,jacb,pada}) and to others
too numerous to name.  It is also well known that quantum mechanics is related to
entropy (and information theory) and we cite here only a recent paper by Caticha
\cite{cath} which summarizes much research in this connection (going back to 
Nagasawa, Nelson, and Smolin among many others - see e.g. \cite{naga,nels,smlo}).
In \cite{c006,c007,c009,caar,caro,cl,crar} we have also described connections
between quantum mechanics, Weyl geometry, and Weyl-Dirac quantum theory
following \cite{absr,audr,agsn,aula,bqcd,cstr,ctmh,dirc,isrt,istr,isrn,nikl,nold,satm,shoj,ssgi,
shgi,whlr} (note that publication information can often be found on the electronic
bulletin board entries which
we often use in our referencing).  The key idea here, going back at least to Audretsch et al and Santamato,
is that quantum mechanics is naturally associated with a Weyl geometry, and hence
to gravitational ideas.  Another manifestation of gravity-quantum connections appears
in papers of Mannheim and Bender (see e.g. \cite{bdmh,mnhm}) where one refers e.g.
to intrinsically quantum mechanical gravity.  This background 
of quantum mechanics in geometry (or gravity) says nothing about
``quantum gravity", as one imagines might be built up in various manners (see e.g.
\cite{kief,klvk,mrkp,orti,rovl,thim,volk,wen}).  Rather the general motivating idea seems
to be in the spirit of accepting a universe guided by information theory plus some ingredients (space, time, matter) and describing how everything fits together.  Never mind
trying to build a universe from scratch.  Of course this encompasses also trying to
understand the microstructure of spacetime, the cosmological constant, and other
matters.

\section{WEYL DIRAC THEORY}	
\renewcommand{\theequation}{2.\arabic{equation}}
\setcounter{equation}{0}

We review first two derivations of Dirac-Weyl theory following \cite{isrt}.  The first
follows Dirac in avoiding Proca terms and the second involves integrable Weyl
geometry.  Thus let $ds^2=g_{ab}dx^adx^b$ be based on a Riemannian metric 
$g_{ab}$ with Levi-Civita connection coefficients (Christoffel symbols)
\bq\label{2.1}
\gG^{\gl}_{\mu\nu}=\frac{1}{2}g^{\gl\gs}\left(\pp_{\nu}g_{\mu\gs}+\pp_{\mu}g_{\nu\gs}
-\pp_{\gs}g_{\mu\nu}\right)
\end{equation}
Recall also that a comma $(,n)$ denotes $\pp_n$ and a semicolon $(;n)$ denotes
a covariant derivative $\na_n$ so that
\bq\label{2.2}
B^{\mu}_{;\nu}= \na_{\nu}B^{\mu}=\pp_{\nu}B^{\mu}+B^{\gs}\gG^{\mu}_{\gs\nu};\,\,
B_{\mu;\nu}=\na_{\nu}B_{\mu}=\pp_{\nu}B_{\mu}-B_{\gs}\gG^{\gs}_{\mu\nu}
\end{equation}
If a vector is parallel transported $x^{\nu}$ to
$x^{\nu}+dx^{\nu}$ one has via (2.2) $({\bf 2A})\,\,dB=0$ where $B=(g_{\mu\nu}B^{\mu}
B^{\nu})^{1/2}$ and one checks that $({\bf 2B})\,\,\na g_{\mu\nu}=0=\na g^{\mu\nu}$.
The (Riemann-Christoffel) curvature tensor is 
\bq\label{2.3}
R^{\gl}_{\gs\mu\nu}=-\pp_{\nu}\gG^{\gl}_{\gs\mu}+\pp_{\mu}\gG^{\gl}_{\gs\nu}-
\gG^{\ga}_{\gs\mu}\gG^{\gl}_{\ga\nu}+\gG^{\ga}_{\gs\nu}\gG^{\gl}_{\ga\mu}
\end{equation}
and if e.g. $B^{\mu}$ is parallel transported around an infinitesimal closed parallelogram
the total change is $({\bf 2C})\,\,\gD B^{\gl}=
B^{\gs}R^{\gl}_{\gs\mu\nu}dx^{\mu}\gd x^{\nu}$ (whereas $\gD B=0$).
\\[3mm]\indent
Now for Weyl geometry both the length and direction of a vector change under
parallel transport and one has a (symmetric) Weyl connection $\hat{\gG}^{\gl}_{\mu\nu}
=\hat{\gG}^{\gl}_{\nu\mu}$ with e.g.
\bq\label{2.4}
dB^{\mu}=-B^{\gs}\hat{\gG}^{\mu}_{\gs\nu}dx^{\nu};\,\,dB=Bw_{\nu}dx^{\nu}
\end{equation}
where $w_{\nu}$ is called a Weyl vector.  In order to be compatible with the a given
metric $g_{ab}$ one must have now 
\bq\label{2.5}
\hat{\gG}^{\gl}_{\mu\nu}=\gG^{\gl}_{\mu\nu}+g_{\mu\nu}w^{\gl}-\gd_{\nu}^{\gl}w_{\mu}
-\gd^{\gl}_{\mu}w_{\nu}
\end{equation}
In particular (denoting Weyl covariant derivatives via a colon $(:)$ 
(as well as a hat) it follows that 
$({\bf 2D})\,\,g^{\mu\nu}_{\,\,:\gl}=\hat{\na}_{\gl}g^{\mu\nu}=-2g^{\mu\nu}w_{\gl}$.
One assumes $\hat{\gG}^{\gl}_{\mu\nu}=\hat{\gG}^{\gl}_{\nu\mu}$ and 
formulas analogous to (2.2) hold for $\hat{\gG}^{\gl}_{\mu\nu}$.  Now however
parallel transport around an infinitesimal closed parallelogram yields $({\bf 2E})\,\,
\gD B=BW_{\mu\nu}dx^{\mu}\gd x^{\nu}$ where $({\bf 2F})\,\,W_{\mu\nu}=\pp_{\nu}w_{\mu}
-\pp_{\mu}w_{\nu}.$  In case $w_{\gl}=\pp_{\gl}\phi$ one has an integrable Weyl 
geometry with $W_{\mu\nu}=0$ and $\gD B=0$.
\\[3mm]\indent
Let now a vector of length B transported around a closed loop to achieve a new
length $({\bf 2G})\,\,B_{new}=B+\int_SBW_{\mu\nu}dS^{\mu\nu}$ with $S$ the area
enclosed by the loop and $dS^{\mu\nu}$ an area element.  In this connection one
considers local Weyl
gauge transformations (WGT) $({\bf 2H})\,\,B\to\tl{B}=exp(\gl)B$ where $\gl(x^{\nu})$ is a 
differentiable function leading to
\bq\label{2.6}
g_{\mu\nu}\to \tl{g}_{\mu\nu}=e^{2\gl}g_{\mu\nu};\,\,g^{\mu\nu}\to\tl{g}^{\mu\nu}=
e^{-2\gl}g^{\mu\nu}
\end{equation}
This is all compatible with Weyl gauge transformations,
namely $({\bf 2I})\,\,w_{\nu}\to\tl{w}_{\nu}=w_{\nu}+\pp_{\nu}\gl$ and 
a Weyl weight (or power) $n=\Pi(\psi)$ is defined for quantities transforming via $({\bf 2J})\,\,\psi\to\tl{\psi}=exp(n\gl)\psi$ (cf. \cite{caht,isrt,rosn} for more discussion).  In particular $({\bf 2K})\,\,
\hat{\gG}^{\gl}_{\mu\nu}\to\hat{\gG}^{\gl}_{\mu\nu}$ with $\Pi(\hat{\gG}^{\gl}_{\mu\nu})=0,
\,\,\Pi(g_{\mu\nu}=2,$ and $\Pi(\sqrt{-g})=4,$ etc.
\\[3mm]\indent
Now in the Dirac formulation of a combined relativity plus electromagnetism (EM)
one arrives at an action (cf. \cite{isrt})
\bq\label{2.7}
I=\int\left[W^{\gl\mu}W_{\gl\mu}-\gb^2R+\gs\gb^2|{\bf w}|^2+(\gs+6)(\na\gb)^2+\right.
\end{equation}
$$\left.+2\gs\gb w^{\gl}\pp_{\gl}\gb+2\gL\gb^4+L_M\right]\sqrt{-g}d^4x$$
Then in order to avoid Proca terms Dirac set $\gs=0$ to arrive at
\bq\label{2.8}
I=\int\left[W^{\gl\mu}W_{\gl\mu}-\gb^2R+6(\na\gb)^2+2\gL\gb^4+L_M\right]\sqrt{-g}d^4x
\end{equation}

\indent
There is however another approach due to Rosen \cite{rosn} via an integrable Weyl-Dirac theory in which the action is given by (2.9) below and $\gs$  can take arbitrary values (cf. \cite{isrt} from which we extract).  Thus take $w_{\nu}=\pp_{\nu}w$ (so $W_{\mu\nu}=0$) and set
$({\bf 2L})\,\,b_{\mu}=c\pp_{\mu}[log(\gb)]$ with $W=w+b$. 
In fact we could now take $w=0$ and still have an integrable Weyl
geometry with $W_{\nu}=c\pp_{\nu}log(\gb)$ (note this is not necessary however). 
The action becomes then
\bq\label{2.9}
I=\int\sqrt{-g}d^4x\left[-\gb^2R+(\gs+6)(\na\gb)^2+2\gL\gb^4+L_M\right]
\end{equation}
(see Remark 2.1 below).
\\[3mm]\indent
In \cite{c009,caar,caro,crar} we considered a conformal map from an Einstein
frame action $({\bf 2M})\,\,S_{GR}=\int d^4x\sqrt{-g}[R-\ga(\na\psi)^2+16\pi L_M]$
to a Brans-Dicke (BD) form called conformal GR, or CGR, which we specify now after deleting the $L_M$ and $\gL$ terms, so that CGR is represented via
\bq\label{2.10}
\hat{S}_{GR}=\int d^4x\sqrt{-g}e^{-\psi}\left[\hat{R}-\left(\ga-\frac{3}{2}\right)(\hat{\na}
\psi)^2\right]=
\end{equation}
$$=\int d^4x\sqrt{-\hat{g}}\left[\hat{\phi}\hat{R}-\left(\ga-\frac{3}{2}\right)\frac{(\hat{\na}
\hat{\phi})^2}{\hat{\phi}}\right]$$
(here $\gO^2=\phi=exp(\psi)=\hat{\phi}^{-1}$).
This is an integrable Weyl geometry based originally on $\hat{g}_{ab}=\gO^2g_{ab},\,\,\gO^2=exp(\psi)=\phi=\hat{\phi}^{-1},\,\,w_{\ga}=\pp_{\ga}\psi=\pp_{\ga}\hat{\phi}/\hat{\phi}$ and a conformal mass $\hat{m}=\hat{\phi}^{-1/2}m$ (cf. \cite{bqcd}). 
Further $\hat{m}$ can be identified
with a quantum mass field ${\mf M}\sim\gb$ from a Bohmian Weyl-Dirac theory with a
quantum potential Q determined via ${\mf M}^2=m^2exp(Q)$ (cf. \cite{bqcd,c006,c007,c009,shoj}).  Thus ${\mf M}=\gb=\hat{m}=\hat{\phi}^{-1/2}m$ and
$\gO^2=\hat{\phi}^{-1}={\mf M}^2/m^2=\gb^2/m^2$.  Now from ({\bf 2L}) for example,
given $w_{\ga}$ already of the form $w_{\ga}=-\pp_{\ga}\psi=\pp_{\ga}\hat{\phi}/\hat{\phi}$,
with $\gb=m\hat{\phi}^{-1/2}$, we find that $w_{\ga}=-2\pp_{\ga}log(\gb)$.  This means
that one could take $w=0$ in $W=w+b$ with $b_{\ga}=c\pp_{\ga}log(\gb)$ in comparing
Weyl-Dirac actions.
\\[3mm]\indent
{\bf REMARK 2.1.}
Note that, given $w_{\ga}=-2\pp_{\ga}log(\gb)$, if we choose $b_{\ga}=c\pp_{\ga}log(\gb)$
with $c\ne 2$, then $W_{\ga}=w_{\ga}+b_{\ga}=(c-2)\pp_{\ga}log(\gb)$ so $W_{\mu\nu}=0$.
However no such manipulation is needed since the unadulterated Wey-Dirac integrand from \cite{isrt} (p. 116) is 
$(-\gb^2\sim \sum (\pp_j\gb)^2\sim(\na \gb)^2\sim (\pp\gb)^2$ and $|{\bf w}|^2=4(\pp\gb)^2/\gb^2$)
\bq\label{2.11}
\left[-\gb^2R+\gs\gb^2|{\bf w}|^2+2\gs\gb w^{\gl}\pp_{\gl}\gb +(\gs+6)(\pp \gb)^2 +
+2\gL\gb^4+L_M\right]
\end{equation}
Here $|{\bf w}|^2=4(\pp\gb/\gb)^2$ so $\gb^2|{\bf w}|^2=4(\pp\gb)^2$ and $\gb w^{\gl}
\pp_{\gl}\gb=-2(\pp\gb)^2$ so, omitting the $\gL$ and $L_M$ terms, the integrand is
\bq\label{2.12}
-\gb^2R+4\gs(\pp\gb)^2-4\gs(\pp\gb)^2+(\gs+6)(\pp\gb)^2=-\gb^2R+(\gs+6)(\pp\gb)^2
\end{equation}
One sees that the $4\gs(\pp\gb)^2$ terms cancel. $\bs$
\\[3mm]\indent
Now, following \cite{c009}, for (2.10) we note first from \cite{c009} that $({\bf 2N})\,\,\hat{R}\hat{\phi}\sqrt{-\hat{g}}=\hat{R}\phi\sqrt{-g}=(\gb^2/m^2)\hat{R}\sqrt{-g}$ and $({\bf 2O})\,\,\gb^2\hat{R}=\gb^2R-6\gb^2\na_{\gl}w^{\gl}+6\gb^2|{\bf w}|^2$.  Hence the first integral in (2.10) takes
the form
\bq\label{2.13}
I_1=\int d^4x\sqrt{-\hat{g}}\hat{\phi}\hat{R}=\int d^4x\sqrt{-g}\phi\hat{R}=\int d^4x\sqrt{-g}\hat{R}\frac{\gb^2}{m^2}=
\end{equation}
$$=\frac{1}{m^2}\int d^4x\sqrt{-g}\left[\gb^2R-6\gb^2\na_{\gl}w^{\gl}+6\gb^2|{\bf w}|^2
\right]$$
and $-\gb^2\na_{\gl}w^{\gl}=-\na_{\gl}(\gb^2w^{\gl})+2\gb\pp_{\gl}w^{\gl}$ so the divergence
term vanishes upon integration and (2.13) becomes
\bq\label{2.14}
I_1=\frac{1}{m^2}\int d^4x\sqrt{-g}\left[\gb^2R+12\gb(\pp_{\gl}\gb)w^{\gl}+6\gb^2|{\bf w}|^2\right]=
\end{equation}
$$=\frac{1}{m^2}\int d^4x\sqrt{-g}\left[\gb^2R+12(\gb\pp_{\gl}\gb)\left(\frac{-2\pp^{\gl}\gb}{\gb}\right)+6\gb^2\left(\frac{2\pp\gb}{\gb}\right)^2\right.+$$
$$\left.+6\gb^2\left(\frac{2\pp\gb}{\gb}\right)^2\right]=\frac{1}{m^2}\int d^4x\sqrt{-g}
\gb^2R$$
The second integral in (2.10) is now [$\gag=\ga-(3/2)$]
\bq\label{2.15}
I_2=-\gag\int d^4x\sqrt{-\hat{g}}\frac{(\pp\hat{\phi})^2}{(\hat{\phi})^2}\hat{\phi}=
\end{equation}
$$=-\gag\int d^4x\sqrt{-g}\phi\left(\frac{-2\pp\gb}{\gb}\right)^2=-\frac{4\gag}{m^2}\int d^4x
\sqrt{-g}(\pp\gb)^2$$
Consequently (2.10) becomes
\bq\label{2.16}
\hat{S}_{GR}=I_1+I_2=\frac{1}{m^2}\int d^4x\sqrt{-g}\left[\gb^2R-4\left(\ga-\frac{3}{2}\right)
(\pp\gb)^2\right]
\end{equation}
This will then represent an integrable Weyl-Dirac theory provided that e.g. $(2.16)\equiv
(2.12)$ which means that (modulo a factor of $(1/m^2))$ we choose $\gs+6=
-4[\ga-(3/2)]$ or $\gs=-4\ga$.
\\[4mm]\indent
{\bf REMARK 2.2}
We note that there is a typo in equation (6.7), p. 59 (and in (3.33), p. 237) in \cite{c009}
namely $3-4\ga$ should be $6-4\ga$.  Also on pp. 58-59 and 236-237 one should have
$\phi=exp(\psi)=\hat{\phi}^{-1}=\gO^2=(\gb/m)^2=({\mf M}/m)^2$ consistently and $L_M$ should be deleted in (6.6)-(6.7), p. 59 and (3.32)-(3.33), p. 237.  A multiplier $\hat{\phi}^2$
should also be added to ({\bf 6C}) on p. 59 and to $(\bgs)$ on p. 237.  In \cite{crar},
on p. 155, line -7, one should change $\psi$ to $-\psi$, and note that the choice of 
comparison actions on p. 157, equation (52), was too constrained; we should have
used the integrable Weyl-Dirac action (2.9) above with arbitrary $\gs$.  
Also in \cite{crar}, p. 157, equation (51) should have $6-4\ga$ and on p. 167, line 13,
insert: $\cdots\,\,u\,\, in\,\, (\bullet\bullet)\,\, denotes\,\, \cdots$.  $\bs$
\\[3mm]\indent
{\bf REMARK 2.3.}
We mention here a related discussion in \cite{nsfo} of quantum mechanics in connection
with Weyl geometry and the Palatini approach to gravitation.  This is connected to conformal aspects of the Palatini approach (as in \cite{accf,cdvi,rstl}).  We mention also
\cite{koch} for a related treatment of geometry and the quantum.  $\bs$

\end{document}